\documentclass[pra,aps,twocolumn,showpacs]{revtex4}
\usepackage{amssymb}
\usepackage{amsmath}
\usepackage[dvips]{graphicx}
\usepackage[english]{babel}

\begin{document}

\title{Cyclic phase in F=2 spinor condensate: Long-range order, kinks, and
roughening transition}
\author{W. V. Pogosov and K. Machida}
\affiliation{Department of Physics, Okayama University, Okayama 700-8530, Japan}
\date{\today }

\begin{abstract}
We study the effect of thermal fluctuations on homogeneous infinite
Bose-Einstein condensate with spin $F=2$ in the cyclic state, when atoms
occupy three hyperfine states with $m_{F}=0, \pm 2$. We use both the
approach of small-amplitude oscillations and mapping of our model on the
sine-Gordon model. We show that thermal fluctuations lead to the existence
of the rough phase in one- and two-dimensional systems, when presence of
kinks is favorable. The structure and energy of a single kink are found. We
also discuss the effect of thermal fluctuations on spin degrees of freedom
in $F=1$ condensate.
\end{abstract}

\pacs{03.75.Kk, 03.75.Lm}
\maketitle

\section{Introduction}

Recent progress in the experimental physics of cold atoms allows for the
creating of quasi-two-dimensional atomic gas either using one-dimensional
optical lattices or applying tight axial trapping \cite%
{Ketterle,Schweikhard,Rychtarik,Smith}. Many properties of low-dimensional
systems differ from that for the three dimensional ones. For instance, it is
well-known that Bose-Einstein condensation is impossible in two-dimensional (%
$2D$) situation at any nonzero temperature, since long wavelength thermal
fluctuations destroy a phase coherence. However, at low temperatures, there
is still a quasi-long-range order, which disappears at the temperature of
Berezinskii-Kosterlitz-Thouless (BKT) transition. Low temperature phase is
often called quasicondensate, and this conception was introduced to the
context of physics of alkali atoms gases in Refs. \cite%
{Shlyapnikov1,Shlyapnikov2}, see also Ref. \cite{Shlyapnikov3}. In was shown
in those works that, in \textit{finite} low-dimensional systems, there is
still a long range order at low temperatures, and one has a true condensate,
but at higher temperatures, coherence is again lost, and we have a
quasicondensate.

The aim of the present paper is to study thermal fluctuations in homogeneous 
\textit{spinor} condensates. We are mainly interested in the cyclic phase of 
$F=2$ condensate, since it has unusual properties like phase locking and
kinks, which are absent in other phases of $F=2$ condensate. In the recent
experiments \cite{Schmal,Chang,Gorlitz,Kuwamoto,Widera}, $F=2$ spinor
Bose-Einstein condensates have been created and studied. The order parameter
in $F=2$ system has five components; and, in the cyclic phase, all the
particles populate three hyperfine states with $m_{F}=0$, $\pm 2$. The
characteristic feature of the cyclic phase is the fact that\ the ground
state energy depends on the relative angle $\chi $ among the phases $S_{j}$
of different components of the order parameter, $\chi =2S_{0}-S_{2}-S_{-2}$,
through the spin-mixing term. This leads to the peculiar phase locking
phenomena, since the energy has a term proportional to $\cos \chi $. In this
paper we concentrate on thermal fluctuations of phases $S_{j}$ of different
components of the order parameter as well as of $\chi $. First, we solve the
Bogoliubov-de Gennes equations and calculate the mean square fluctuations of 
$S_{j}$. Each individual phase $S_{j}$ in this case behaves in the way
similar to the case of scalar system. Namely, there is no long-range order
for $S_{j}$ in one and two dimensions, whereas this order is kept for $3D$
situation. At the same time, the long-range order is still preserved for $%
\chi $. After this, we note that the Bogoliubov-de Gennes equations can be
insufficient for the analysis of thermal fluctuations in the cyclic state,
since this approach uses an expansion of the energy in the vicinity of one
of the infinite number of equivalent minima at $\chi =2\pi l$, $l$ being an
integer number, and does not take into account a global periodic structure
of the energy in functional space. In other words, $\cos \chi $ contribution
to the energy is changed by the quadratic potential well with the \textit{%
infinite} height. One of the possible solutions of the Gross-Pitaevskii
equations for the cyclic phase is a \textit{kink}, which separates two
spatial domains with $\chi $ different by $2\pi $ from each other. At zero
temperature, kink is energetically unfavorable. However, at finite
temperature formation of kinks can become favorable due to entropic reasons.
We determine analytically the structure and energy of a single kink. Also we
note that our model can be mapped on the well-known sine-Gordon model. For
this model, it was established before that, in low dimensions, a so called
roughening transition can occur, when the system becomes unlocked from one
of the minima of $\chi $. In $1D$ situation, a long range order for $\chi $
is absent for any nonzero temperature due to thermally-excited kinks, and
the system is rough. If the system has a finite length, a nonzero
temperature appears, below which still there is a long range order for $\chi 
$. We estimate this temperature from simple entropic arguments. In two
dimensions, a roughening transition occurs at some finite nonzero
temperature. From known results for the sine-Gordon model and some simple
qualitative considerations we show that the temperature of the roughening
transition is close to the BKT critical temperature and possibly coincides
with it. We also discuss briefly the case of ferromagnetic $F=1$ condensate.
Based on the solutions of the Bogoliubov-de Gennes equations, we find that
there is no long-range order in the one- and two-dimensional systems for the
direction of spin. In finite systems, the order still exists at low
temperatures, similarly to the phase coherence in scalar condensates.

The paper is organized as follows. In Section II we solve the Bogoliubov-de
Gennes equations for $F=2$ cyclic phase and find correlation functions. In
Section III we determine the structure and energy of a single kink. In
Section IV we discuss the order-disorder transition due to the proliferation
of kinks. In Section V we study the case of ferromagnetic $F=1$ condensate.
We conclude in Section VI.

\section{Cyclic phase: small amplitude oscillations}

We consider infinite homogeneous $F=2$ condensate with a given density of
particles $n$ in zero magnetic field. The energy of the system depends on
three interaction parameters $\alpha ,$ $\beta ,$ and $\gamma ,$ which can
be defined as \cite{Ciobanu,Ueda1}%
\begin{eqnarray}
&&\alpha =\frac{1}{7}(4g_{2}+3g_{4}),  \label{1} \\
&&\beta =-\frac{1}{7}(g_{2}-g_{4}),  \label{2} \\
&&\gamma =\frac{1}{5}(g_{0}-g_{4})-\frac{2}{7}(g_{2}-g_{4}),  \label{3}
\end{eqnarray}%
where ($q=0,2,4$) 
\begin{equation}
g_{q}=\frac{4\pi \hbar ^{2}}{m}a_{q}  \label{4}
\end{equation}%
and $a_{q}$ is the scattering lengths characterizing collisions between
atoms with the total spin $0$, $2$, and $4$. In real atomic condensates, $%
\alpha \gg \beta ,$ $\gamma $.

The order parameter in $F=2$ case has five components $\Psi _{i}$ $%
(i=-2,-1,0,1,2)$. The energy of the system is given by \cite{Machida,Ho}

\begin{eqnarray}
F &=&\int d\mathbf{r}\left[ -\frac{\hbar ^{2}}{2m}\Psi _{j}^{\ast }\Delta
\Psi _{j}+\frac{\alpha }{2}\Psi _{j}^{\ast }\Psi _{k}^{\ast }\Psi _{j}\Psi
_{k}\right.  \notag \\
&&+\frac{\beta }{2}\Psi _{j}^{\ast }\Psi _{l}^{\ast
}(F_{a})_{jk}(F_{a})_{lm}\Psi _{k}\Psi _{m}  \notag \\
&&\left. +\frac{\gamma }{2}\Psi _{j}^{\ast }\Psi _{k}^{\ast }\Psi _{-j}\Psi
_{-k}(-1)^{j}(-1)^{k}\right] ,  \label{5}
\end{eqnarray}%
where integration is performed over the system volume, repeated indices are
summed, $F_{a}$ $(a=x,y,z)$ is the angular momentum operator, which can be
expressed in a usual matrix form.

In the absence of magnetic field and rotation, condensate can be in three
different states \cite{Ciobanu}, as seen from Eq.~(5). These states are
called ferromagnetic, cyclic and polar \cite{Ciobanu}. In cyclic phase, $%
\Psi _{\pm 1}=0$ and $\Psi _{-2}=\frac{\sqrt{n}}{2}e^{i\theta },$ $\Psi _{0}=%
\frac{\sqrt{n}}{\sqrt{2}},$ $\Psi _{2}=-\frac{\sqrt{n}}{2}e^{-i\theta },$
where $\theta $ is an arbitrary phase (energy of the system is degenerate
with respect to $\theta $). The ground state value of $\chi $ in these
notations is zero. Depending on values of scattering lengths $a_{q}$
ferromagnetic, cyclic or polar phase has the lowest energy \cite{Ciobanu}.
Extended Gross-Pitaevskii equations can be obtained as usually from the
condition of minimum of free energy of the system Eq.~(5). In the cyclic
state, the last term in the right-hand side of Eq. (5) contains a
spin-mixing contribution:

\begin{equation}
F_{sm}=\gamma \left\vert \Psi _{0}\right\vert ^{2}\left\vert \Psi
_{2}\right\vert \left\vert \Psi _{-2}\right\vert \cos \chi .  \label{6}
\end{equation}

Now we analyze small-amplitude oscillations of the order parameter in the
cyclic phase. The fluctuations of phase of the given component of the order
parameter can be expressed through the fluctuations of the order parameter
itself as $\delta S_{j}=Im(\delta \Psi _{j})/\left\vert \Psi _{j}\right\vert 
$. For \ $\delta \chi $, we have $\delta \chi =2\delta S_{0}-\delta
S_{2}-\delta S_{-2}$.

The deviations of the five components of the order parameter from their
equilibrium values $\delta \Psi _{j}$ can be represented as

\begin{eqnarray}
\delta \Psi _{j} &=&\sum\limits_{\mathbf{k}}\left\{ c_{\mathbf{k}}^{(j)}\exp
(i\mathbf{kr}+i\omega _{\mathbf{k}}t)\right.  \notag \\
&&\left. +d_{\mathbf{k}}^{(j)\ast }\exp (-i\mathbf{kr}-i\omega _{\mathbf{k}%
}t)\right\} ,  \label{7}
\end{eqnarray}%
where $c_{\mathbf{k}}^{(j)}$ and $d_{\mathbf{k}}^{(j)}$ are constants to be
found. Fluctuations of $\Psi _{\pm 1}$ are decoupled from that for $\Psi
_{0} $, $\Psi _{\pm 2}$, as follows for the Bogoliubov-de Gennes equations
for the cyclic phase. The spectrum for oscillations of $\Psi _{0}$, $\Psi
_{\pm 2}$ has three branches. For the first time, it was obtained in Ref. 
\cite{Suom}, and our results coincide with those results: 
\begin{eqnarray}
\hbar \omega _{\mathbf{k}}^{(1)} &=&\frac{\hbar ^{2}\mathbf{k}^{2}}{2m}%
+2\gamma n,  \label{8} \\
\hbar \omega _{\mathbf{k}}^{(2)} &=&\sqrt{\left( \frac{\hbar ^{2}\mathbf{k}%
^{2}}{2m}\right) ^{2}+2\alpha n\frac{\hbar ^{2}\mathbf{k}^{2}}{2m}},
\label{9} \\
\hbar \omega _{\mathbf{k}}^{(3)} &=&\sqrt{\left( \frac{\hbar ^{2}\mathbf{k}%
^{2}}{2m}\right) ^{2}+4\beta n\frac{\hbar ^{2}\mathbf{k}^{2}}{2m}}.
\label{10}
\end{eqnarray}%
The thermal distribution of quasiparticles is given by ($j=1$, $2$, $3$) 
\begin{equation}
N_{\mathbf{k}}^{(j)}=\frac{1}{\exp (\hbar \omega _{\mathbf{k}%
}^{(j)}/k_{B}T)-1}.  \label{11}
\end{equation}%
One can also find eigenvectors corresponding to the branches (8)-(10).
Eigenvectors for Eqs. (9) and (10) give no contribution to $\delta \chi $,
as it can be seen from the solutions of the Bogoliubov-de Gennes equations.
This is the first branch (8), which is responsible for the fluctuations of $%
\chi $. Physically, this is due to the fact that only branch (8) depends on
the value of $\gamma $, which yields the potential well for $\chi $ via the
spin-mixing term (6). Finally, we obtain simple expression for $\delta \chi $%
: 
\begin{eqnarray}
\delta \chi &=&\frac{1}{2i\sqrt{V}\sqrt{n}}\sum\limits_{\mathbf{k}}\left\{
\exp (i\mathbf{kr}+i\omega _{\mathbf{k}}t)\right.  \notag \\
&&\left. -\exp (-i\mathbf{kr}-i\omega _{\mathbf{k}}t)\right\} .  \label{12}
\end{eqnarray}

Next we can determine the behavior of the mean square fluctuations of $\chi $
at large distances. We found that this quantity in all dimensions tends to a
constant, when the distance tends to infinity. This means that there is a
long-range order for $\chi $. The reason is the presence of the gap in the
excitation energy (8). As an example, we present below the derivation and
results for the $2D$ situation. From Eq. (12) we have\bigskip 
\begin{equation}
\left( \delta \chi (r)-\delta \chi (0)\right) ^{2}=-\frac{1}{4Vn}%
\sum\limits_{\mathbf{k}_{1},\mathbf{k}_{2}}A_{\mathbf{k}_{1}}A_{\mathbf{k}%
_{2}},  \label{13}
\end{equation}%
where

\begin{eqnarray*}
A_{\mathbf{k}} &=&\exp (i\omega _{\mathbf{k}}^{(1)}t)\left( \exp (i\mathbf{%
kr)-}1\right) \\
&&-\exp (-i\omega _{\mathbf{k}}^{(1)}t)\left( \exp (-i\mathbf{kr)-}1\right) .
\end{eqnarray*}%
After averaging over the time, only terms with $k_{1}=k_{2}$ survive in the
expansion (13) and for the correlator we have

\begin{eqnarray}
\left\langle \left[ \delta \chi (\mathbf{r})-\delta \chi (0)\right]
^{2}\right\rangle _{T} &=&-\frac{1}{8\pi Vn}\sum\limits_{k,\varphi
_{1},\varphi _{2}}N_{\mathbf{k}}^{(j)}\times  \notag \\
&&\left( e^{ikr\cos \varphi _{1}}-1\right) \left( e^{ikr\cos \varphi
_{2}}-1\right) ,  \label{14}
\end{eqnarray}%
where $\varphi _{1}$ and $\varphi _{2}$ are angles between $\mathbf{r}$ and $%
\mathbf{k}_{1}$ and $\mathbf{k}_{2}$, respectively. Now we can switch from
the integration to the summation, and after the integration over $\varphi
_{1}$ and $\varphi _{2}$ we get

\begin{equation}
\left\langle \left[ \delta \chi (\mathbf{r})-\delta \chi (0)\right]
^{2}\right\rangle _{T}=\frac{1}{2\pi n}\int kdkN_{\mathbf{k}}^{(1)}\left(
1-J_{0}(kr)\right) ^{2},  \label{15}
\end{equation}%
where $J_{0}(r)$ is the Bessel function. This integral can be evaluated
analytically at large $r$ and in the limits of low and high temperatures. At
low temperatures, $k_{B}T\ll \gamma n$, $N_{\mathbf{k}}^{(1)}=\exp (-\hbar
\omega _{\mathbf{k}}^{(j)}/k_{B}T)$. At high temperatures, $k_{B}T\gg \gamma
n$, $N_{\mathbf{k}}^{(1)}=k_{B}T/\hbar \omega _{\mathbf{k}}^{(j)}$. Finally,
we have

\begin{widetext}
\begin{equation}
\left\langle \left[ \delta \chi (\mathbf{r})-\delta \chi (0)\right]
^{2}\right\rangle _{T}=\left\{ 
\begin{array}{ll}
T/T_{d}\exp (-2\gamma n/k_{B}T), & k_{B}T\ll \gamma n, \\ 
T/T_{d}\ln (k_{B}T/2\gamma n), & k_{B}T\gg \gamma n,%
\end{array}%
\right.   \label{16}
\end{equation}
\end{widetext}

where $T_{d}=2\pi \hbar ^{2}n/k_{B}m$ is the temperature of quantum
degeneracy.

The second and third branches of the spectrum, given by Eqs. (9) and (10),
are responsible for the fluctuations of phases of individual components of
the order parameter, $S_{0}$, $S_{\pm 2}$. Our calculations of the mean
square fluctuations of these phases revealed the behavior, which is similar
to the scalar condensate. The long-range order is absent in one- and
two-dimensional situations. It is interesting to note that, at the same
time, the long-range order exists for the linear combination of these
phases, $\chi $. Fluctuations of populations of different hyperfine states
are small at $T$ $\ll T_{d}$.

\section{Kink: structure and energy}

In the previous Section we have studied small-amplitude oscillations in the
vicinity of homogeneous solution to the extended Gross-Pitaevskii equations.
However, these equations have also a nonhomogeneous solution corresponding
to the kink or domain wall between two spatial regions with values of $\chi $
different by $2\pi l$ from each other, since the energy of the system is
degenerate with respect to $l$. Kinks are known to play an important role in
the physics of low-dimensional systems, see e.g. the textbook \cite{Chaikin}%
. Now we find the structure and energy of the one-dimensional kink in the
cyclic phase for the most important case, when $l=1$. We assume that all the
quantities depend only on one coordinate $x$. For $\chi $ we have a boundary
conditions $\chi (\infty )=2\pi $, $\chi (-\infty )=0$. The phase-dependent
part of the energy consists of two contributions, one is the kinetic energy,
and another one is the spin-mixing term (6):

\begin{widetext}
\begin{equation}
F_{ph}=\int_{-\infty }^{+\infty }dx\left[ \frac{\hbar ^{2}}{2m}%
\sum_{j=-1}^{1}\left\vert \Psi _{2j}\right\vert ^{2}\left( \nabla
S_{2j}\right) ^{2}+\gamma \left\vert \Psi _{0}\right\vert ^{2}\left\vert
\Psi _{2}\right\vert \left\vert \Psi _{-2}\right\vert \cos \chi \right] .
\label{17}
\end{equation}

We can rewrite this expression in the diagonal form as

\begin{equation}
F_{ph}=\int_{-\infty }^{+\infty }dx\left[ \frac{\hbar ^{2}n}{8m}\frac{1}{4}%
\left( \nabla \chi \right) ^{2}+\frac{\gamma n^{2}}{4}\cos \chi +\frac{\hbar
^{2}n}{8m}\left( \frac{1}{2}\left( \nabla S_{2}-\nabla S_{-2}\right) ^{2}+%
\frac{1}{4}\left( \nabla S_{2}+\nabla S_{-2}+2\nabla S_{0}\right)
^{2}\right) \right] .  \label{18}
\end{equation}%
\end{widetext}

One can easily see from Eq. (18) that the minimum energy solution for the
kink corresponds to the conditions $\nabla S_{2}-\nabla S_{-2}=0$ and $%
\nabla S_{2}+\nabla S_{-2}+2\nabla S_{0}=0$. In this case, the energy (18)
can be mapped on the sine-Gordon energy functional. The function $\chi (x)$
is found from the condition of minimum of $F_{ph}$: $\delta F_{ph}/\delta
\chi =0$ and boundary conditions. Finally, we have

\begin{eqnarray}
\chi (x) &=&2\pi -4\tan ^{-1}\exp (-x/\xi _{\chi }),  \label{19} \\
S_{2}(x) &=&S_{-2}(x)=-S_{0}(x)=-\frac{1}{4}\chi (x),  \label{20}
\end{eqnarray}%
where%
\begin{equation}
\xi _{\chi }=\sqrt{\frac{\hbar ^{2}}{2m}\frac{1}{4\gamma n}}  \label{21}
\end{equation}%
is the healing length for $\chi $.\ It also gives a characteristic length of
the kink. The spatial structure of a single kink is presented in Fig. 1,
where we have plotted $x$-dependences of $\chi $, $S_{\pm 2}$, and $S_{0}$.
A single kink moving in the space is a \textit{soliton}. As seen from Eq.
(18), the energy of the kink $F_{kink}$ scales as $\gamma \xi _{\chi }\sim 
\sqrt{\gamma }$. More accurate calculation based on Eq. (19) yields%
\begin{equation}
F_{kink}=n^{3/2}\sqrt{\frac{\hbar ^{2}\gamma }{m}}.  \label{22}
\end{equation}%
Note that in real atomic condensates $\gamma $ has to be much smaller than $%
\alpha $ and, therefore, $\xi _{\chi }$ should far exceed the coherence
length, which has a meaning of a length-scale for the density modulations.

The same structure of the kink can be obtained from the extended
Gross-Pitaevskii equations supplemented by the boundary conditions for $\chi 
$. These equations are rather cumbersome and we do not present them here
(see e.g. our previous work \cite{Pogos}). It is easy to see by the direct
substitution that Eqs. (19)-(21) with the \textit{constant} populations of
each magnetic sublevel yield the \textit{exact} solution to these equations.

\section{Roughening}

At zero temperature, the sine-Gordon system in any dimension is locked in
one of equivalent minima corresponding to $\chi =2\pi l$. At nonzero
temperature, presence of kinks can be favorable due to their entropic
contribution to the free energy \cite{Chaikin}. Kinks destroy the long-range
order and lead to the rough phase, in which height of fluctuations diverges
with tending the system size to infinity. In one dimensional system of
length $L$ ($L\gg \xi _{\chi }$), the entropy corresponding to the single
kink can be estimated as $\ln L/\xi _{\chi }$, where $L/\xi _{\chi }$ is
just a number of places to put a kink. Therefore, a roughening temperature is%
\begin{equation}
T_{R}^{1D}\approx \frac{F_{kink}}{k_{B}\ln L/\xi _{\chi }}.  \label{23}
\end{equation}%
In the case of infinite $1D$ system ($L\rightarrow \infty $), $T_{R}^{1D}=0$%
; and this result is very different from the result of small-amplitude
oscillations approach, which shows a presence of long-range order for $\chi $%
. Relation (23) is well-known for the sine-Gordon model and can be derived
using more rigorous analysis, see e.g. Ref. \cite{Ares} and references
therein. Note that in $1D$ finite system, $T_{R}^{1D}\rightarrow 0$, as $%
\gamma \rightarrow 0$.

In two dimensions, a roughening temperature $T_{R}^{2D}$ can be also
estimated in a simple manner. In this case, $F_{kink}$ given by Eq. (22) is
the energy of a unit length of a kink. The 'critical nucleus' for the
formation of the rough phase is represented by the step for $\chi $\ of
height $2\pi $\ in a functional space with a perimeter $\approx 2\pi \xi
_{\chi }$. The total energy of such a nucleus is of the order of $2\pi
F_{kink}\xi _{\chi }=\pi \hbar ^{2}n_{s}/2m$, where we have changed density
of atoms $n$ by superfluid density $n_{s}$ just below the roughening
transition. It is interesting to note that this energy is independent on $%
\gamma $. A roughening temperature $T_{R}^{2D}$ can be estimated by equating
the nucleus energy to the thermal one, $T_{R}^{2D}\approx \pi \hbar
^{2}n_{s}/2k_{B}m$. A more rigorous calculation for the $2D$ sine-Gordon
model, based on renormalization group analysis \cite{Chaikin}, yields the
same result:%
\begin{equation}
T_{R}^{2D}=\frac{\pi \hbar ^{2}n_{s}}{2k_{B}m},  \label{24}
\end{equation}%
which is independent on $\gamma $. The expression (24) for the roughening
temperature coincides with the well-known result for the temperature of the
BKT transition, see e.g. \cite{Shlyapnikov3}. To describe properly the
behavior of the system in this strongly fluctuative region at high
temperatures, one needs more careful and detailed analysis. However, from
the considerations presented here, we can conclude that a long range order
for $\chi $ in $2D$ systems survives up to quite high temperatures (of the
order of the BKT critical temperature). Possibly, the roughening transition
occurs simultaneously with the BKT transition, and both types of topological
defects, kinks and vortices, proliferate together.

If the size of $1D$ or $2D$ finite system is less than $\xi _{\chi }$ (which
can be much larger than the coherence length), a formation of a kink is
impossible, and, therefore, we expect that $\chi $ is nearly constant inside
the cloud. At the same time, a value of $\chi $ can be different from $2\pi
l $ due to thermal fluctuations. We have studied the similar situation
before in Ref. \cite{Pogosov}, where the case of harmonically trapped
quasi-two-dimensional $F=1$ condensate was treated. An infinite homogeneous $%
F=1$ condensate in zero magnetic field can be either in polar or
ferromagnetic states, where atoms populate only one or two hyperfine states
and the spin-mixing terms in the energy are zero in the equilibrium.
However, in the case of a trapped rotated condensate containing vortices,
there are some regions on the phase diagram, where all the three hyperfine
states are populated, and the energy depends on the relative phase among
them \cite{Mizushima,Isoshima,Mizushima1,Mizushima2,Kita,Isoshima1}. This
vortex phases can be both locally and globally stable, and some of them,
like Mermin-Ho vortex, have an axial symmetry, which makes them rather
simple. Another method to create such states was recently used
experimentally in Ref. \cite{Chapman}, where a microwave energy was injected
to the system leading to the redistribution of the particles from spin $-1$
state to spin $0$ and $1$ states. We see that the cyclic phase in $F=2$
condensate is one more example of the systems, where spin-mixing is
important. It is remarkable that, in this case, spin-mixing is happening in
the ground state and even in absence of rotation, applied magnetic fields
and other external perturbations, as in $F=1$ case. For the spin-mixing
dynamics in $F=1$ condensate, see Ref. \cite{Pu}. Note that recently, a new
experimental method for nondestructive study of internal degrees of freedom
of spinor condensates was proposed in Ref. \cite{Higbie}.

\section{Ferromagnetic state in $F=1$ spinor condensate}

In this Section we discuss thermal fluctuations in spinor $F=1$ condensate.
We concentrate on the ferromagnetic state. The energy of the system is
independent on the direction of the spin but it depends on the gradients of
spin. The similar situation exists for the phase of the order parameter in
scalar condensate (the energy is independent on its value, but depends on
the gradient), and therefore we can expect the similar behavior, namely,
absence of long-range order in low dimensions. This result was obtained
before for the large class of discrete and continuous spin models \cite%
{Chaikin}. We assume that in the ground state all the atoms occupy the only
one hyperfine state, $\Psi _{-1}=\Psi _{0}=0$, $\Psi _{1}=\sqrt{n}$, and
study small-amplitude oscillations induced by the temperature. In the ground
state, spin is oriented along the $z$-direction, $S_{z}=1$, $S_{x}=S_{y}=0$.
Perturbations of $\Psi _{0}$ lead to that of $S_{x}$ and $S_{y}$:%
\begin{equation}
\delta S_{x}=\frac{1}{\sqrt{2n}}\left( \delta \Psi _{0}+\delta \Psi
_{0}^{\ast }\right) ,  \label{25}
\end{equation}%
\begin{equation}
\delta S_{y}=\frac{i}{\sqrt{2n}}\left( \delta \Psi _{0}-\delta \Psi
_{0}^{\ast }\right) .  \label{26}
\end{equation}

The spectrum for the $F=1$ condensate was obtained in Refs. \cite{Machida,Ho}%
. The frequency of the mode, responsible for fluctuations of $\Psi _{0}$, is
given by

\begin{equation}
\hbar \omega =\frac{\hbar ^{2}\mathbf{k}^{2}}{2m}.  \label{27}
\end{equation}%
Using Eqs. (25)-(27) we have calculated mean square fluctuations for $S_{x}$
and $S_{y}$ at large distances and found an absence of long-range order in $%
1D$ and $2D$ cases. As an example, we show here the correlator in the $2D$
situation:%
\begin{equation}
\left\langle \left[ S_{x}(\mathbf{r})-S_{x}(0)\right] ^{2}\right\rangle _{T}=%
\frac{T}{T_{d}}\ln \frac{r}{\lambda _{T}},  \label{28}
\end{equation}%
where

\begin{equation}
\lambda _{T}=\frac{\hbar }{\sqrt{2mk_{B}T}}.  \label{29}
\end{equation}%
The derivation is similar to the case of cyclic phase in $F=2$ condensate,
presented in Section IV. The result for $S_{y}$ is the same. For $1D$ case,
the correlators behave as $\sim r$. This implies that in low-dimensional
systems there is no long-range order in the direction of spin. However, the $%
2D$ system can be divided into blocks of a characteristic size $L$, $\xi
_{s}\ll L\ll \lambda _{T}\exp (T_{d}/T)$, with nearly the same direction of
spin in each block ($\xi _{s}\sim \sqrt{\frac{\hbar ^{2}}{2m}\frac{1}{\beta n%
}}$ is the spin healing length). Different blocks are uncorrelated with each
other. At temperatures of the order of $2T_{d}/\ln (k_{B}T_{d}/\beta n)$, $%
\xi _{s}$ becomes comparable to $\lambda _{T}\exp (T_{d}/T)$. In this case,
the size of each block becomes of the order of $\xi _{s}$.

In the quasi-$1D$ and $2D$ trapped condensates, which have finite sizes
exceeding $\xi _{s}$, at low temperatures, one can expect the same
orientation of spins throughout the system, and at higher temperatures, the
cloud should consist of uncorrelated blocks. This is similar to the problem
of fluctuating phase in quasi-$1D$ and $2D$ scalar condensates \cite%
{Shlyapnikov1,Shlyapnikov2,Shlyapnikov3}. Note that the case of
ferromagnetic $F=2$ condensate is similar to that of spin-1 system.

\section{Conclusions}

We have studied the effect of thermal fluctuations on homogeneous infinite
Bose-Einstein condensate with spin $F=2$. We were interested in the cyclic
state of this system, in which all the particles occupy three hyperfine
states with $m_{F}=0$, $\pm 2$, and the energy depends on the relative phase 
$\chi =2S_{0}-S_{2}-S_{-2}$ through the spin-mixing term. Using
Bogoliubov-de Gennes equations, we have calculated mean square fluctuations
of $S_{0}$, $S_{\pm 2}$ and found absence of long-range order in one- and
two-dimensional situations, but presence of this order for $\chi $. Then, we
went beyond the small-amplitude oscillations approach and mapped our problem
on the sine-Gordon model. A structure and energy of a single kink separating
two topological sectors with different values of $\chi $ were found. In $1D$
and $2D$ situations, thermal proliferation of kinks can lead to the
roughening transition destroying a long-range order for $\chi $.
One-dimensional infinite system is always in a rough phase, and finite $1D$
system, as well as infinite $2D$ system, experience roughening transition at
finite temperatures, whose values we have estimated. At the end, we have
also discussed thermal fluctuations in ferromagnetic $F=1$ system and found
an absence of long-range order in the direction of spin in low-dimensional
situations, as it can be expected from the Mermin-Wagner-Berezinskii
theorem. We defined the typical size of each block, within which still there
is an order in spin degrees of freedom. We expect that finite systems should
be in the ordered phase at low temperatures.

\acknowledgments

Authors acknowledge very useful discussions with H. Adachi and T. K. Ghosh.
W. V. Pogosov is supported by the Japan Society for the Promotion of Science.

\section{Figure captions}

Fig. 1. The structure of a single kink in the cyclic state. Solid curve
denotes $\chi $, dashed one corresponds to $S_{0}$, and dotted curve shows $%
S_{2}=S_{-2}$.


\begin{references}

\bibitem{Ketterle}A. Gorlitz, J. M. Vogels, A. E. Leanhardt, C. Raman, 
T. L. Gustavson, J. R. Abo-Shaeer,  A. P. Chikkatur, S. Gupta, S. Inouye, 
T. Rosenband, and W. Ketterle, Phys. Rev. Lett. {\bf 87}, 130402 (2001).

\bibitem{Schweikhard}V. Schweikhard, I. Coddington, P. Engels, 
V. P. Mogendorff, and E. A. Cornell, Phys. Rev. Lett. {\bf 92}, 040404 (2004). 

\bibitem{Rychtarik}D. Rychtarik, B. Engeser, H.-C. Nagerl, and R. Grimm, 
Phys. Rev. Lett. {\bf 92}, 173003 (2004). 

\bibitem{Smith}N. L. Smith,  W. H. Heathcote, G. Hechenblaikner, 
E. Nugent, and C. J. Foot, J. of Phys. B At. Mol. and Opt. 
Phys. {\bf 38}, 223 (2005).

\bibitem{Shlyapnikov1}D. S. Petrov, M. Holzmann, and G. V. Shlyapnikov, 
Phys. Rev. Lett. {\bf 84}, 2551 (2000).

\bibitem{Shlyapnikov2}D. S. Petrov, G. V. Shlyapnikov, and J. T. M. Walraven,
Phys. Rev. Lett. {\bf 87}, 050404 (2001).

\bibitem{Shlyapnikov3}D. S. Petrov, D. M. Gangardt, and G. V. Shlyapnikov,
J. Phys. IV France {\bf 1} (2005); cond-mat/0409230.

\bibitem{Schmal}H. Schmaljohann, M. Erhard, J. Kronj\"{a}ger, M. Kottke, S. van Staa,
L. Cacciapuoti, J. J. Arlt, K. Bongs, and K. Sengstock,
Phys. Rev. Lett. {\bf 92}, 040402 (2004).

\bibitem{Chang}M.-S. Chang, C. D. Hamley, M. D. Barrett, J. A. Sauer, K. M. Fortier,
W. Zhang, L. You, and M. S. Chapman,
Phys. Rev. Lett. {\bf 92}, 140403 (2004).

\bibitem{Gorlitz}A. G\"{o}rlitz, T. L. Gustavson, A. E. Leanhardt, R. L\"{o}w, A. P. 
Chikkatur, S. Gupta, S. Inouye, D. E. Pritchard, and W. Ketterle,
Phys. Rev. Lett. {\bf 90}, 090401 (2003).

\bibitem{Kuwamoto}T. Kuwamoto, K. Araki, T. Eno, and T. Hirano,
Phys. Rev. A {\bf 69}, 063604 (2004).

\bibitem{Widera}A. Widera, F. Gerbier, S. Folling, T. Gericke, O. Mandel, and I. Bloch,
cond-mat/0604038.

\bibitem{Ciobanu}C.  V. Ciobanu, S.-K. Yip, and Tin-Lun Ho,
Phys. Rev. A {\bf 61}, 033607 (2000).

\bibitem{Ueda1}M. Koashi and M. Ueda,
Phys. Rev. Lett. {\bf 84}, 1066 (2000).

\bibitem{Machida} T. Ohmi and K. Machida, J. Phys. Soc. Jpn. {\bf 67}, 1822 (1998). 

\bibitem{Ho} T.- L. Ho, Phys. Rev. Lett. {\bf 81}, 742 (1998).

\bibitem{Suom}J. P. Martikainen and K. A. Suominen, J. of Phys. B At. Mol. and Opt. 
Phys. {\bf 34}, 4091 (2001).

\bibitem{Chaikin}P. M. Chaikin and T. C. Lubensky, {\it Principles of condensed
matter physics}, Cambridge Univ. Press, Cambridge (1995).

\bibitem{Pogos}W. V. Pogosov, R. Kawate, T. Mizushima, and K. Machida, Phys. Rev. A {\bf 72}, 
063605 (2005). 

\bibitem{Ares}S. Ares, J. A. Cuesta, A. Sanchez, and R. Toral, 
Phys. Rev. E {\bf 67}, 046108 (2003).

\bibitem{Pogosov}W. V. Pogosov and K. Machida,
cond-mat/0602119.

\bibitem{Mizushima}T. Mizushima, K. Machida, and T. Kita, Phys. Rev. Lett. {\bf 89}, 030401 (2002).

\bibitem{Isoshima}T. Isoshima and K. Machida, Phys. Rev. A {\bf 66},  023602 (2002). 

\bibitem{Mizushima1}T. Mizushima, K. Machida, and T. Kita, Phys. Rev. A {\bf 66}, 053610 (2002).

\bibitem{Mizushima2}T. Mizushima, N. Kobayashi, and K. Machida, Phys. Rev. A {\bf 70},  043613 (2004).

\bibitem{Kita}T. Kita, T. Mizushima, and K. Machida, Phys. Rev. A {\bf 66}, 061601 (2002).

\bibitem{Isoshima1}T. Isoshima, K. Machida, and T. Ohmi, J. Phys. Soc. Jpn. {\bf 70}, 1604 (2001).

\bibitem{Chapman}M. S. Chang, Q. Qin, W. Zhang, L. You, and M. S. Chapman, 
Nature Physics {\bf 1}, 111 (2005). 

\bibitem{Pu}H. Pu, C. K. Law, S. Raghavan, J. H. Eberly, and N. P. Bigelow, 
Phys. Rev. A {\bf 60}, 1463 (1999).

\bibitem{Higbie}J. M. Higbie, L. E. Sadler, S. Inouye, A. P. Chikkatur, S. R. Leslie, 
K. L. Moore, V. Savalli, and D. M. Stamper-Kurn, Phys. Rev. Lett. {\bf 95}, 050401 (2005).

\end{references}
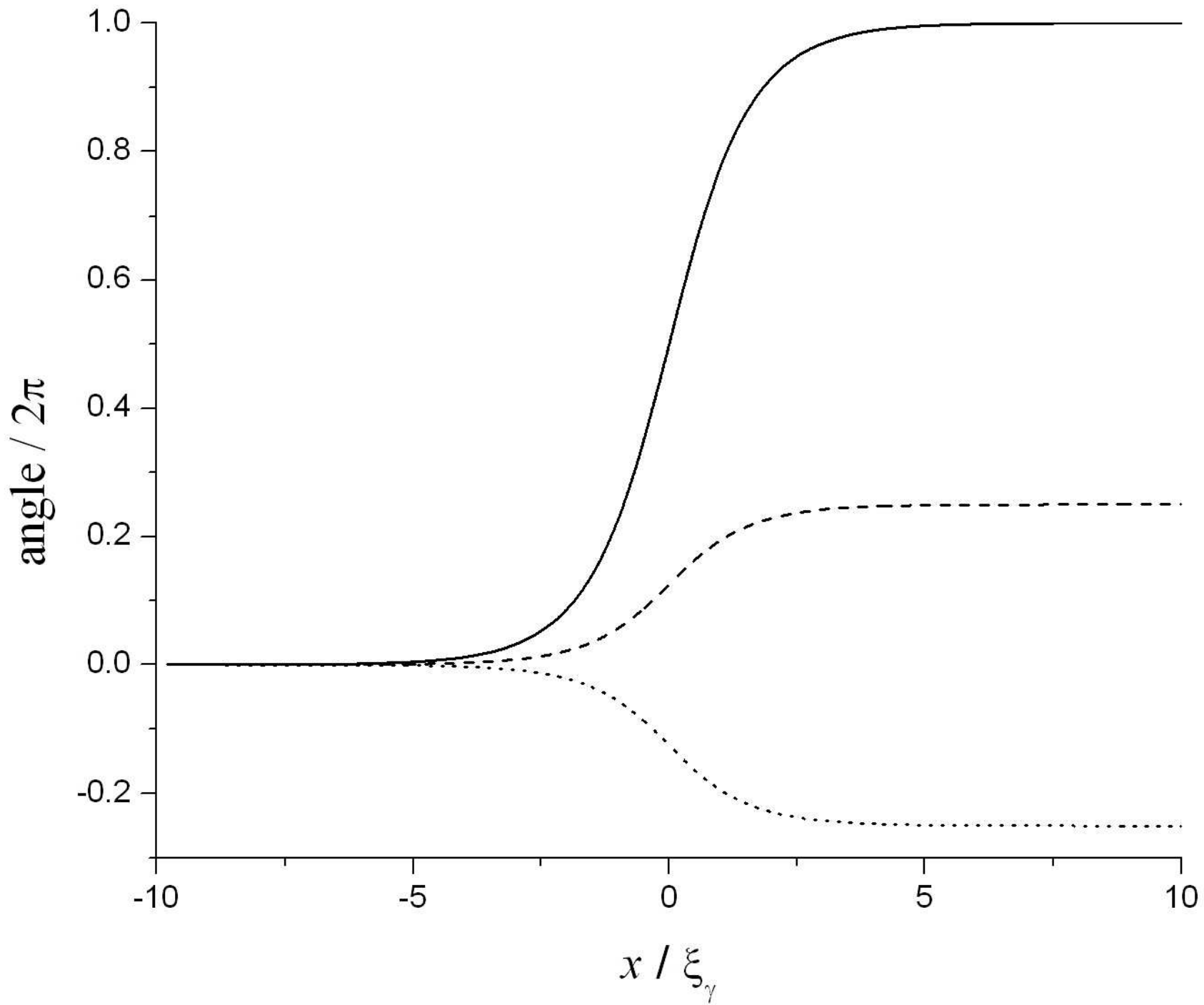
\end{document}